\documentclass[prl,twocolumn,showpacs,superscriptaddress]{revtex4}

\usepackage{graphicx,bm,amssymb,amsmath}
\begin{document}

\title{Force mobilization and generalized isostaticity in jammed packings of frictional grains}

\author{Kostya Shundyak} \affiliation{Instituut--Lorentz,
Universiteit Leiden, Postbus 9506, 2300 RA Leiden, The
Netherlands}

\author{Martin van Hecke} \affiliation{Kamerlingh Onnes Lab,
Universiteit Leiden, Postbus 9504, 2300 RA Leiden, The
Netherlands}

\author{Wim van Saarloos} \affiliation{Instituut--Lorentz,
Universiteit Leiden, Postbus 9506, 2300 RA Leiden, The
Netherlands}

\date{\today}

\begin{abstract}
We show that in slowly generated 2$d$ packings of frictional
spheres, a significant fraction of the friction forces lies at the
Coulomb threshold --- for small pressure $p$ and friction
coefficient $\mu$, about half of the contacts. Interpreting these
contacts as  constrained leads to a generalized concept of
isostaticity, which relates the maximal fraction of fully
mobilized contacts and contact number. For $p\to 0$, our
frictional packings approximately satisfy this relation over the
full range of $\mu$. This is in agreement with a previous
conjecture that gently built packings should be marginal solids at
jamming. In addition, the contact numbers and packing densities
 scale with both $p$ and $\mu$.
\end{abstract}

 \maketitle

Models of {\em frictionless} polydisperse particles with
finite-range repulsive forces exhibit a well-defined ``jamming
point'' J in the limit that the confining pressure $p$ goes to
zero \cite{jamming_nature,epitome}. In the vicinity of J on the
jammed side, i.e. for $p\agt 0$, the average contact number,
packing density, elastic constants, vibrational modes and response
functions all show scaling behavior as a function of pressure
\cite{epitome,wyart,wouter_rollover}. This scaling is intimately
connected to the fact that when point J is approached by preparing
packings at lower and lower pressures, such packings become {\em
isostatic}: a simple constraint counting argument for hard spheres
in $d$ dimensions yields that for $p\to0$,  the average number of
contacts per interacting particle $z$ equals the fictionless
isostatic value $z^0_{\rm iso}=2d$ \cite{moukarzel,wittencheck}.

The picture which is emerging for {\em frictional} packings is
much more diffuse, since there are now two control parameters ($p$
and $\mu$), and more importantly, packing densities and contacts
numbers depend on the preparation method and history. This is
because the Coulomb condition for the frictional force is an
inequality: it specifies, for each static contact, that the
tangential force $f_{\rm t} $ be less than or equal to the
friction coefficient $\mu$ times the normal force $f_{\rm n}$:
$|f_{\rm t}| \le \mu f_{\rm n}$. If in view of this inequality we
treat these tangential forces as {\em independent} new degrees of
freedom in the constraint counting, the isostatic value jumps from
$z^0_{\rm iso}=2d $ to $z^\mu_{\rm iso}=d+1$, and in $d$ dimension
frictional packings for $p\to0$ can in principle occur for {\em
any} $z$ in the range $z_{\rm iso}^\mu \equiv d+1 \le z \le
z^0_{\rm iso}$ \cite{counting}.

In practice, however, for a given experimental \cite{swinney} or
numerical \cite{unger,silbert_geo,makse_fric} protocol some
reproducible value $z$ is found. The sudden jump of the isostatic
contact number with $\mu$ is not reflected in a jump of $z_{\rm
J}(\mu) \equiv z(\mu,p\to 0)$: numerically, $z_{\rm J}(\mu)$ is
found to vary smoothly from $z^0_{\rm iso}$ at small $\mu$ to some
limiting value at large $\mu$ \cite{unger}. The large $\mu$ limit
may or may not coincide with $z^\mu_{\rm iso}$, and $z$ is
generally smaller and closer to the isostatic value the slower the
packing is prepared \cite{makse_fric}.

\begin{figure}[tb]
\includegraphics*[width=7.6cm]{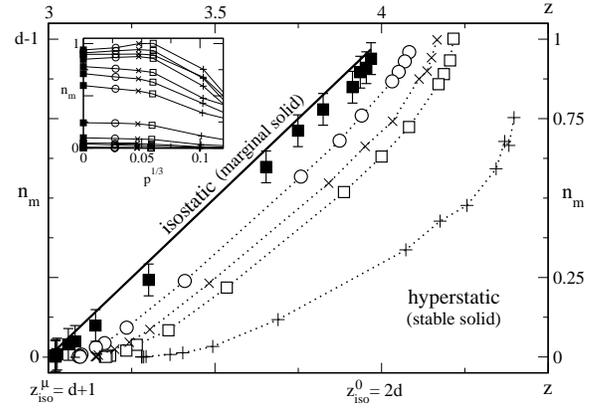}
\caption{\label{generalfig} Relation between the number of fully
mobilized forces per particle $n_m$ and contact number $z$. The
full line indicates the maximum of $n_m$ and such packings are
marginal, while below this line one finds hyperstatic stable
packings. The data points refer to numerically obtained values of
$n_m$ in 2 dimensions,
 for $p \sim 2\times10^{-4}$ (+), $p \sim 5\times10^{-5}$
 ($\square$), $p \sim 2\times10^{-5}$ ($\times$) and
 $p \sim 5\times10^{-6}$
 ($\circ$).
$n_m$ and $z$ behave smoothly as function of $p^{1/3}$, and by
extrapolation we obtain our $p=0$ estimate indicated by the black
squares (see inset and main text).}\end{figure}

As stressed by Silbert {\em et al.} \cite{silbert_geo} and
Bouchaud \cite{bouchaud_leshouches}, there is a natural way in
which the discontinuity in the isostatic contact numbers is not
reflected in $z_{\rm J}(\mu)$, which hinges on the notion of
maximizing the number of fully mobilized or ``plastic'' contacts,
i.e., those at the Coulomb failure threshold for which $m=1$,
where $m\equiv |f_{\rm t}|/(\mu f_{\rm n})$
\cite{silbert_geo,bouchaud_leshouches}. Since at fully mobilized
contacts  tangential and normal forces are related, this leads to
additional constraints in the counting arguments:
Introducing $n_{\rm m}$ as the number of fully mobilized contacts
per particle in a packing with $N_{\rm i} $ interacting particles,
the $zdN_{\rm i}/2$ force degrees of freedom should be larger than
the total number of constraints provided by the $N_{\rm
i}d(d+1)/2$ force and torque balance equations \cite{counting} and
the $n_{\rm m}N_{\rm i}$ mobilization constraints. This gives:
\begin{equation}
n_{\rm m} \le z- z^\mu_{\rm iso}.
\end{equation}
From this point of view, packings with $ n_{\rm m} = z- z^\mu_{\rm
iso}$ are in fact isostatic or marginal, while packings with
$n_{\rm m} < z- z^\mu_{\rm iso}  $ are hyperstatic (see
Fig.~\ref{generalfig}).

In this paper, we will show that gently prepared packings support
this scenario over a surprisingly wide range of friction
coefficients. The distribution function $P(m)$ of such packings
indeed naturally splits up in a peak at $m=1$ and a broad flat
part for $m<1$ (Fig.~2), and these packings actually tend to be
marginal at jamming, i.e., to lie close to this generalized
isostaticity line in Fig.~\ref{generalfig}. The picture that
emerges is that if we prepare the packings sufficiently slowly,
they get stuck in a marginal state. Such a marginal scenario also
occurs in, e.g.,  spinglasses \cite{bouchaud_leshouches}, charge
density waves \cite{coppersmith87} and phase organization
\cite{tang87}).

The fact that our well-equilibrated packings approach a
well-defined limit opens up the possibility to study the
asymptotic scaling behavior as a function of pressure and friction
coefficient $\mu$. We have therefore also investigated the effect
of the applying pressure on repeatedly and gently created packings
over a whole range of friction coefficients, and find that contact
numbers $z$ and packing densities $\phi$ of the packings do
exhibit scaling with $p$ and $\mu$. The scaling of $\phi$ and $z$
with $p$ are related to the form of the interparticle potential
and consistent with previous findings for the frictionless case.
The scaling of $z$ and $\phi$ with $\mu$ appear to be independent
of the force law --- we have at present no good physical
understanding of this scaling.

{\em Model and simulation method ---} We numerically build 2$d$
packings of $N_{\rm p}=1000$ polydisperse spheres that interact
through 3$d$ Hertz-Mindlin forces or through
one-sided-linear-springs-plus-friction \cite{HMfoot} in a square
box with periodic boundary conditions. The data reported below are
all for the 3$d$ Hertz-Mindlin forces. Following \cite{ellak} our
units are such that the mass density, the average particle
diameter and the Young's modulus of the grains are 1. The Poisson
ratio of the grains is taken to be zero, and there is no gravity.
As in \cite{ellak} the packings are constructed by cooling an
initial low density state where the particles have a small
velocity, while slowly inflating the particle radii by multiplying
them with a common scale factor $r_s$. This factor is determined
by solving the damped equation $r_s'' = - 4 \omega_0 r_s' -
\omega_0^2 \left[p(t,r_s)/p-1 \right]r_s$, where $ \omega_0 \sim
6*10^{-2}$, $p(t,r_s)$ is the instant value of the pressure and
$p$ the target pressure. This ensures a very gentle equilibration
of the packings.
In our analysis of forces and contact numbers, we always take out
rattlers, by considering contact forces less than $10^{-3}$ times
the average force broken and removing particles with less than two
contacts. For each packing, we determine the total number of
contacts $N_{\rm c}$ and the total number of interacting particles
$N_{\rm i}$ (the total number of particles minus the rattlers) ---
$z\equiv 2N_{\rm c}/N_{\rm i}$. For each value of $p$ and $\mu
\in[10^{-3},10^3]$, 30 realizations have been constructed with a
polydispersity of 20\%. We occasionally checked that taking 60
realizations, a different polydispersity or different damping
parameters lead to similar results. In comparison with other
simulations where the particles settled under gravity
\cite{silbert_geo} or were quenched rapidly \cite{makse_fric}, our
algorithm prepares the packings  more gently, in the sense  that
it results in low packing densities and coordination numbers.

\begin{figure}[tb]
\includegraphics*[width=8.3cm]{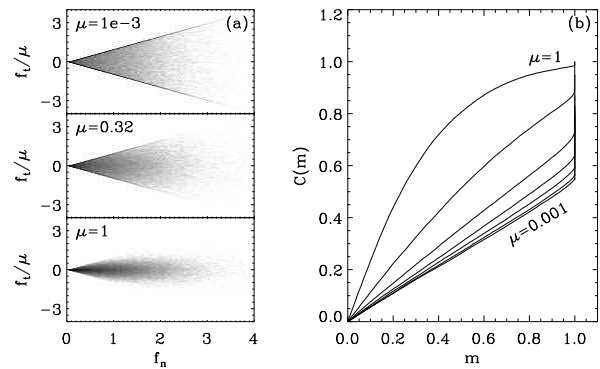}
\caption{\label{mobfn} Mobilization at  $p=2\times10^{-5}$. (a)
Scatter plots of $f_{\rm t}/\mu$ versus $f_{\rm n}$ for three
packings at  $\mu=0.001, 0.32 $ and $1$. The probability density
of normalized tangential $(f_t/\mu)$ and normal $(f_n)$ forces
exhibits a singularity on the Coulomb cone for small $\mu$, which
rapidly diminishes for larger $\mu$ (all forces are normalized so
that $\langle f_n \rangle =\!1$). (b) The cumulative distribution
of the mobilization $C(m) \equiv \int^m dm' P(m')$ exhibits a
clear jump near $m=1$ --- data shown here is for
$\mu=10^0,10^{-0.5},10^{-1},\dots,10^{-3}$. }
\end{figure}

{\em The density $n_{\rm m}$ of fully mobilized contacts ---}
 The joint probability distribution of the normal and
frictional contact forces clearly show that for small $\mu$, a
substantial amount of forces lie on the Coulomb cone, i.e., have
$m=1$, while for larger $\mu$ the fraction of fully mobilized
contacts diminishes  (Fig.~2a). A priori it would appear to be
difficult to determine numerically whether a contact is fully
mobilized with $m=1$ or elastic (non-mobilized) with $m<1$, but as
Fig.~2b shows, the cumulative distribution $C(m) \equiv \int^m
dm^\prime P(m^\prime)$ exhibits a clear jump at $m\!=\!1$. The
value of $n_{\rm m}$ equals $z/2\left[1\!-\!C(m\!\to\!1)\right]$,
and we find that for small friction about half of the contacts
(one contact per particle) is at the Coulomb treshold! Especially
for small $\mu$, $C(m)$ is linear in $m$, which means that the
distribution of non-mobilized forces is flat --- in other words,
non-mobilized contacts are not biased towards higher contact
numbers.

Our estimates for $n_m$ and $z$ for $p\to0$ and a range of $\mu$
lie very close to  the generalized isostaticity line
(Fig.~\ref{generalfig}). Note that we have extrapolated contact
numbers and $n_m$ to estimate the zero pressure limit (see the
inset of Fig.~1 and Fig.~3). The close proximity of $n_m$ and $z$
to the marginal line presents, to our knowledge, the strongest
support to date for the marginal solid scenario described above:
when frictional packings are sufficiently gently prepared, they
form a marginally stable jammed solid which in a generalized sense
is an isostatic solid. We expect that the deviations from the
generalized isostaticity will be larger the faster the granular
particles are compressed or quenched; earlier simulations already
give indications for this \cite{silbert_geo,makse_fric}.

\begin{figure}[tb]
\includegraphics*[width=8.6cm]{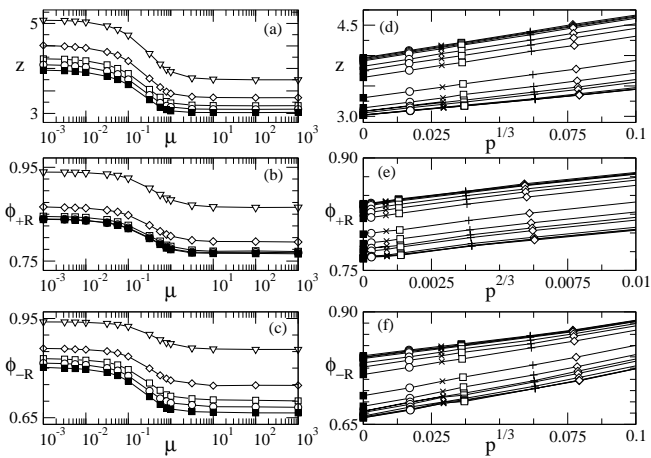}
\caption{\label{phiz} Variation of contact numbers $z$ and packing
density $\phi$ as function of pressure $p$ and friction
coefficient $\mu$. Errorbars are smaller then the symbol size.
(a-c) The variation of the contact number $z$, the packing density
including rattlers $\phi_{\rm +R}$, and the packing density
excluding rattlers $\phi_{\rm -R}$ as a function of $\mu$. Symbols
indicate data at pressures $p \sim 4\times10^{-3} (\triangledown),
5\times10^{-4} (\diamond), 2\times10^{-4} (+), 5\times10^{-5}
(\square), 2\times10^{-5} (\times), 5\times10^{-6} (\circ)$. Based
on the extrapolation illustrated in panels (d-f), we also show the
estimated values at $p=0$ ($\blacksquare$). Even though $\phi_{\rm
+R}$ and $\phi_{\rm -R}$ differ substantially, their variation
with $\mu$ appears very similar. (d-f) $z$ scales as $p^{1/3}$ and
$\phi_{\rm +R}$ as $p^{2/3}$, which allows us to extrapolate to
zero pressure. Surprisingly, the packing density $\phi_{\rm -R}$
does not scale convincingly with $p^{2/3}$, but rather as
$p^{1/3}$. Symbols are as in panel a-c.}
\end{figure}

{\em Scaling behavior of $z$ and $\phi$ ---} Since our packings
for small $p$ approach the generalized isostaticity line, one may
wonder how contact number and packing density $\phi$ change when
moving away or along this line. Since the number of rattlers is
strongly dependent on the pressure $p$ and on the friction
coefficient $\mu$, we have found it illuminating to study both the
density with the rattlers excluded and included, $\phi_{\rm -R}$
and $\phi_{\rm +R}$, respectively. Note that for small pressure
and small friction about 4\% of the particles are rattlers, which
rises to 12\% for large values of the friction. The results of our
analysis are shown in Fig.~3a-c. As a function of $\mu$, the
overall variation of $z$ in Fig.~3a  is very similar to results
obtained by contact dynamics \cite{unger}, and again the density
variations in Fig. 3bc mimic that of $z$. As a function of $p$,
our data is consistent with the scaling relation
$z(\mu,p)-z(\mu,0)\sim p^{1/3}$ (Fig.~3d). This allows us to
extrapolate with confidence to zero pressures, giving $z(\mu \ll
1,0)=3.98\pm0.02$ and $z(\mu \gg 1,0)=3.00\pm 0.02$, which are
close to the frictionless and frictional isostatic bounds,
$z^0_{\rm iso}=4$ and $z^\infty_{\rm iso}=3$, respectively. For
the whole range of $\mu$ we find that the change in density
including rattlers scales as $\phi_{\rm +R}(\mu,p)-\phi_{\rm
+R}(\mu,0)\sim p^{2/3}$  (Fig.~\ref{phiz}e). This is consistent
with the scaling of the density in frictionless packings upon
compressing a given packing \cite{epitome}, and with the variation
$K\sim ({\rm d} \phi_{\rm +R}/{\rm d}p)^{-1}\sim p^{1/3}$ of the
compression modulus $K$ with pressure \cite{epitome,wouter}.
Interestingly, the density excluding rattlers, $\phi_{\rm -R}$
appears to vary instead as $p^{1/3}$ (Fig.~3f).

For our Hertz-Mindling forces, the $p^{1/3}$ scaling for $z$ is
consistent with the scaling $z-z^0_{\rm iso} \sim \sqrt{\delta}$
observed also for frictionless particles \cite{epitome,wouter},
where $\delta$ is the typical dimensionless overlap of the
particles. We have checked that our results do only trivially
depend on the details of the force law: for one-sided harmonic
springs the $z$ and $\phi$ scale as function of $p^{1/2}$ (not
shown). The fact that $z$ scales with $p$ similarly as for
frictionless systems was seen in some studies \cite{makse_fric}
but not in others \cite{silbert_geo}. Both the presence of this
scaling and fact that our packings reach the generalized
isostaticity line for $p\to 0$ may be related to our very slow
rate of equilibration.

From the zero pressure extrapolations discussed above, we can
study the variation of the contact number and densities at
jamming. The results of this analysis are summarized in Fig.~4,
with details given in the figure caption. In particular we find
$z(\mu,0)$ to decrease for small $\mu$ as  $\mu^{0.7\pm 0.1}$.
That indeed $z$ decreases rapidly with $\mu$ is also clear from
the 3$d$ data of \cite{silbert_geo}, which appear to fit a
powerlaw behavior $\Delta z\sim \mu^{0.5}$ reasonably well.
Whether the density changes for small $\mu$ with a nontrivial
exponent different from 1 is less clear from our data. We can not
draw any firm conclusion from our data regarding the functional
$\mu$-dependence for large friction but the variation of contact
number with density appears is consistent with an exponent of 1.7.
Similar scalings are obtained for linear instead of Hertzian
contact laws.

\begin{figure}[tb]
\includegraphics*[width=8.6cm]{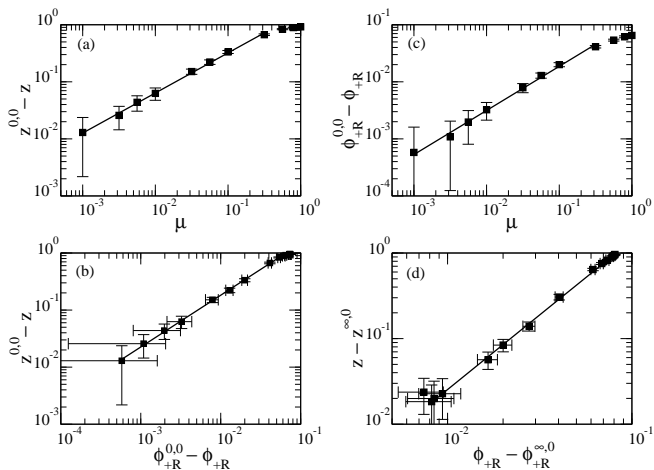}
\caption{Scaling of the zero pressure, extrapolated, contact
numbers and packing densities with the friction coefficient $\mu$.
The extrapolated values at zero (infinite) friction are labelled
as $0,0$ ($0,\infty$). (a-c) When $\mu \to 0$ and $p\to 0$, $z$
approaches $z^{0,0} \approx$ 3.975[20], while $\phi_{\rm +R}$
approaches $\phi_{\rm +R}^{0,0} \approx$ 0.8395[5]. For finite but
small $\mu$, $z$ and $\phi_{\rm +R}$ appear to scale as: (a)
$(z^{0,0}-z) \sim \mu^{0.70[10]}$ and (b) $(\phi_{\rm
+R}^{0,0}-\phi_{\rm +R}) \sim \mu^{0.77[10]}$. (c) The contact
number and packing deviate similarly from this scaling when $\mu $
approaches one, and so $(z^{0,0}-z) \sim (\phi_{\rm
+R}^{0,0}-\phi_{\rm +R})^{0.91[10]}$. (d) In the limit of infinite
friction and zero pressure, $z$ approaches $z^{\infty,0}=
3.00$[2], while $\phi_{\rm +R}$ approaches $\phi^{\infty,0}_{\rm
+R} = 0.758$ [10]. The deviations from this limiting values also
appear to be related by a scaling relation of the form
$(z-z^{\infty,0}) \sim (\phi_{\rm +R} -\phi^{\infty,0}_{\rm +R}
)^{1.7\left[2\right]}$.}
\end{figure}

{\em Summary and Outlook ---} Our results substantiate the
scenario that when a packing is gently prepared, it gets jammed in
a (near) marginal state, where enough contacts get stuck at the
Coulomb failure threshold to make the packing a marginal solid.
Note that this is different from what engineers refer to as
``incipient failure everywhere'' --- the idea that one can deal
with the Coulomb inequality by turning it into an equality for
{\em all} contacts \cite{degennes}. Our results here show that
this overestimates the number of fully mobilized contacts. Our
results suggest a lower boundary for the contact number, and
possibly for the packing densities too, that can be obtained for
finite $\mu$, whereas naive counting would suggest that
$d$-dimensional packings with any contact number between $d+1$ and
$2d$ could arise.

An immediate implication of our results is that the response
properties of such gently prepared packings will have a strong
tendency to show nonlinear response, depending very sensitively on
the behavior of the plastic contacts: if these remain fixed at the
Coulomb threshold, the fact that these packings are near
isostaticity will give many low-frequency modes and will make
these packings very soft. If these contacts yield, however,
irreversibility effects will dominate.

The contact numbers and densities that characterize gently
prepared packings show various nontrivial scaling relations as a
function of $\mu$ and $p$. The scaling of $z\sim p^{1/3}$ and
$\phi_{+R} \sim p^{2/3}$ with $p$ are similar to those found for
frictionless Hertzian packings - but these scalings seem to work
equally well over the whole range of $\mu$. The scaling of
$\phi_{-R}$ is more puzzling. It is very well possible that the
asymptotic behavior for very small $p$ crosses over to the
familiar $p^{2/3}$ behavior, but we can not access this regime  at
present. In addition, for 3d packings the fraction of rattlers may
be smaller than for 2d, so that there we expect less of this
effect. Nevertheless, the question whether one should include or
exclude rattlers is subtle --- see also \cite{grr}.

The scaling of $z$ and $\phi_{+R}$ with $\mu$ is new and presently
not understood, but may give indirect evidence for strong
correlations between the tangential forces. Suppose we think of
the tangential forces $f_{\rm t}$ as small randomly pointing
perturbations of the net forces on the particles for $\mu \ll 1$.
In a domain of linear scale $L$, these tangential forces add up to
a total force of order $\mu f_{\rm n}L^{d/2}$. This is comparable
to the normal force scale $f_{\rm n}$ on a scale $L_\mu \simeq
\mu^{-2/d}$. It might therefore be natural to suppose that on this
scale the tangential forces allow to reduce $z$ by replacing a
single contact. Since $\Delta z L_\mu^d={\mathcal O}(1)$, this
would suggest $\Delta z \sim \mu^2$, in strong contrast to the
data.

{\em Acknowledgement} We are grateful to Ell\'ak Somfai for use of
his numerical routines and to Wouter Ellenbroek, Leo Silbert and
Corey O' Hern for illuminating discussions. KS acknowledges
financial support from the FOM foundation and MvH support from
NWO/VIDI.


\begin{thebibliography}{99}

\bibitem{jamming_nature}
A. J. Liu and S. Nagel, Nature {\bf 396}, 21 (1998).

\bibitem{epitome}
C. S. O'Hern, S.A. Langer, A. J. Liu and S. R. Nagel, Phys. Rev.
Lett. {\bf 88}, 075507 (2002); C.S. O'Hern, L.E. Silbert, A.J. Liu
and S.R. Nagel, Phys. Rev. E {\bf 68}, 011306 (2003).

\bibitem{wyart}
M. Wyart, S.R. Nagel, T.A. Witten, Euro. Phys. Letters, {\bf 72},
486-492, (2005); M. Wyart, L.E. Silbert, S.R. Nagel, T.A. Witten,
 Phys. Rev. E {\bf72} 051306 (2005); M. Wyart, Ann Phys {\bf30}, (3) 1
 (2005).

\bibitem{wouter_rollover}
W.G. Ellenbroek {\em et al.}, in {\em Powders and Grains} edited
by R. Garc\'ia-Rojo {\em et al.} (A.A. Balkema, Rotterdam,
 377 (2005); W. G. Ellenbroek, E. Somfai, M. van Hecke, and W. van Saarloos,
cond-mat/0604157.

\bibitem{moukarzel} C.F. Moukarzel, Phys. Rev. Lett.
{\bf 81}, 1634 (1998).

\bibitem{wittencheck}  A.V. Tkachenko and T.A. Witten,
Phys. Rev. E {\bf 60}, 687 (1999).

\bibitem{counting} For a packing of $N_{\rm i}$ interacting
particles (non-rattlers), there are $dN_{\rm i}$ force balance
equations and $d(d-1)N_{\rm i}/2$ torque balance equations, and
the number of forces is $zN_{\rm i}/2$. If all tangential forces
are arbitrary, this gives, $z\ge d+1$. Together with the $zN_{\rm
i}/2$ constraints that all interacting particles just touch as
$p\to 0$, we get $d+1\le z\le 2d$ at jamming.

\bibitem{swinney} M. Schr\"oter, D. I. Goldman, and H. L. Swinney,
Phys. Rev. E {\bf 71}, 030301(R) (2005).

\bibitem{unger} T. Unger, J. Kert\'esz, and D. E. Wolf, Phys. Rev. Lett. {\bf 94}, 178001 (2005).

\bibitem{silbert_geo}
L. E. Silbert, D. Ertas, G. S. Grest, T. C. Halsey, and D. Levine,
Phys. Rev. E \textbf{65}, 031304 (2002).

\bibitem{makse_fric}
H. A. Makse, N. Gland, D. L. Johnson and L. Schwartz, Phys. Rev. E
{\bf 70}, 061302 (2004); H. P. Zhang and H. A. Makse, Phys. Rev. E
\textbf{72}, 011301 (2005).

\bibitem{bouchaud_leshouches}
J.-P. Bouchaud, \textit{in Slow Relaxations and nonequilibrium
dynamics in condensed matter Liquids, Freezing and Glass
Transition, Les Houches Session LXXVII}, edited by J.-L. Barrat,
M. Feigelman, J. Kurchan, J. Dalibard (Springer Berlin/Heidelberg,
2004).

\bibitem{coppersmith87} S. N. Coppersmith and P. B. Littlewood,
Phys. Rev. B {\bf 36}, 311 (1987).

\bibitem{tang87} C. Tang, K. Wiesenfeld, P. Bak, S. Coppersmith, and P. Littlewood
Phys. Rev. Lett. {\bf 58}, 1161 (1987).

\bibitem{HMfoot}
I.e., normal force $f_{\rm n} \sim \delta^{\alpha-1}$ with $\delta
$ the overlap between particles, $\alpha=5/2$ (2) for
Hertz-Mindlin (Linear spring) forces, and tangential force
increment ${\rm d} f_{\rm t} \sim \delta^{\alpha-2} {\rm d}t$ with
${\rm d}t$ the relative tangential displacement change, provided
$f_{\rm t}\leq \mu f_{\rm n}$.

\bibitem{ellak}
E. Somfai, J.-N. Roux, J. H. Snoeijer, M. van Hecke and W. van
Saarloos, Phys. Rev. E {\bf 72}, 021301 (2005).

\bibitem{wouter} W. Ellenbroek, E. Somfai, M. van Hecke and W. van Saarloos, cond-mat/0604157.

\bibitem{degennes} P. G. de Gennes, Rev. Mod. Phys. {\bf 71}, 374
(1999).

\bibitem{grr} L. E. Silbert, A. J. Liu and S. R. Nagel, Phys. Rev.
E {\bf 73}, 041304 (2006).

\end{thebibliography}
\end{document}